\documentstyle[twoside,fleqn,espcrc2]{article}
\input epsf

\newcommand{\AmS}{{\protect\the\textfont2
  A\kern-.1667em\lower.5ex\hbox{M}\kern-.125emS}}

\title{ Deflation of Eigenvalues for GMRES in Lattice QCD}

\author{Ronald B. Morgan\address{Department of Mathematics, Baylor 
								University, Waco, TX 76798-7328} and 
        Walter Wilcox\address{Department of Physics, Baylor 
								University, Waco, TX 76798-7316}}
       
\begin{document}

\begin{abstract}
Versions of GMRES with deflation of eigenvalues are applied to lattice 
QCD problems. Approximate eigenvectors corresponding to the smallest 
eigenvalues are generated at the same time that linear equations are 
solved. The eigenvectors improve convergence for 
the linear equations, and they help solve other right-hand sides. 
\end{abstract}

\maketitle

\section{Introduction}

This paper looks at the iterative solution of complex, non-Hermitian systems
of linear equations associated with a lattice QCD problem in particle
physics. Let the $n$ by $n$ system of equations be $Ax=b$. We use a
version of the GMRES method that deflates eigenvalues and can solve systems
with multiple right-hand sides. We give examples that show deflating
eigenvalues can make a significant improvement in the convergence for QCD
matrices. 

We work with the non-Hermitian Wilson-Dirac matrix
$M(x,y)$ given by,
\begin{eqnarray}
M(x,y) = I - \kappa \sum_\mu [(1+\gamma_\mu) U_\mu^\dagger (x-a_\mu)
\nonumber \\ \delta_{x,y+a_\mu} + (1-\gamma_\mu) U_\mu (x) \delta_{x,y-a_\mu} ], 
\end{eqnarray}
because we perfer to consider systems for which multiple-mass
shifts are possible. Our lattice here is of size $12^3 
\times 24$. After the standard even/odd preconditioning,
the lattice QCD problem has the form,
\begin{equation}A_e x_e = b_e + \kappa D_{eo} b_o, \ \ 
A_e = I - \kappa^2 D_{eo} D_{oe}.
\end{equation}
where $A_e$ is of dimension 248,832. We set $\kappa = 0.1575$, which for 
this size matrix and beta value (6.0) means working essentially at 
$\kappa_{critical}$. We consider a
typical, non-exceptional configuration, $U_\mu$. 
The method is designed to efficiently solve the
multiple right-hand sides of $A x=b$ that occur when
the \lq \lq all to all" propagators used for disconnected diagrams are
calculated. We will use multiple right-hand sides formed with Z(2) noise
vectors. The matrix is not extremely sparse, with about 200 nonzeros per
row. 

Krylov subspace methods are iterative methods for solving large
systems of linear equations. With approximate solution $x_0$, and
residual vector $r_0 = b - A x_0$, the linear equations can be recast as
$A(x-x_0)=r_0.$  Then the Krylov subspace of dimension
$m$ for this problem is
$Span\{r_0, Ar_0, A^2 r_0, ... , A^{m-1}r_0 \}.$
The conjugate gradient method is for Hermitian problems, and 
GMRES~\cite{SaSc,Frommer} is a well-known Krylov method for the 
non-Hermitian case. 

The effectiveness of Krylov methods is often controlled by the distribution
of eigenvalues. The existence of small eigenvalues can significantly slow 
the convergence rate. This motivates attempts to remove or deflate some 
eigenvalues from the effective spectrum for an iterative method.

\section{The Method}

The GMRES method fully orthogonalizes a basis for the Krylov subspace. As 
the iteration proceeds, the expense and storage both grow. This
makes occasional restarting necessary. We refer to each pass
through the GMRES iterations between restarts as a ``cycle''.

Several approaches have been proposed for deflating eigenvalues from GMRES; 
see~\cite{GMRES-DR} for references. For deflation in QCD problems, see
\cite{dF,EdHeNa,DoLeLiZh,NeEiLiNeSc}. Our approach differs from previous
QCD deflation in that the eigenvalue problem is solved
simultaneously with the linear equations. 

In this paragraph, we describe the method for the first right-hand side.
After the first GMRES cycle, approximate eigenvectors called harmonic
Ritz vectors, are computed from the same Krylov subspace generated
for solving the linear equations. Let the approximate eigenvectors
corresponding to the $k$ approximate eigenvalues of smallest modulus be 
$\tilde y_1, \tilde y_2, \ldots \tilde y_k$. For the next cycle, the
subspace used for GMRES is 
\begin{eqnarray}
Span\{r_0, A r_0, A^2 r_0, A^3 r_0, \ldots 
,A^{m-k-1} r_0,  \nonumber \\  \tilde y_1, \tilde y_2, \ldots 
\tilde y_k \}. 
\end{eqnarray} 
Having the approximate eigenvectors in the subspace accomplishes two
things. The corresponding eigenvalues are deflated once the approximate
eigenvectors are moderately accurate. Also the approximate eigenvectors
are improved as the method proceeds. It sometimes takes several cycles to
improve the approximate eigenvectors to the point that they are useful. 
This method is called GMRES-DR (GMRES with 
deflated restarting)~\cite{GMRES-DR}. GMRES-DR gives 
the approximate eigenvectors in the form of a short Arnoldi-type recurrence 
\begin{equation}
AV_k = V_{k+1} \bar H_k \label{shortar},
\end{equation} 
where $V_k$ is an $n$ by $k$ orthonormal matrix whos columns span the
subspace of approximate eigenvectors, and $\bar H_k$ is a full $k+1$ by $k$
matrix. 

Now we look at solving the second and subsequent right-hand sides. We use 
the eigenvector information that was generated in solving the first 
right-hand side. An efficient approach in which the eigenvalues are deflated  
outside of the GMRES cycles with a simple projection is
called GMRES-Proj~\cite{GMRES-DR}. With the short Arnoldi-type 
recurrence~(\ref{shortar}) from GMRES-DR, we need to store only $k+1$
vectors of length $n$ in order to have access to both the approximate
eigenvectors and their products with $A$. This allows for fairly inexpensive
projections. We use a 
minimum residual projection~\cite{GMRES-DR} here. The GMRES-Proj method 
applies cycles of standard GMRES, with a projection over 
the approximate eigenvector subspace in between cycles. This
projection is not needed between all of the GMRES cycles.

\section{Experiments}

We test the approaches just discussed with the matrix $A_e$, given earlier. 

\begin{figure}
\begin{center}
\leavevmode
\epsfxsize=2.8 in
\epsfbox{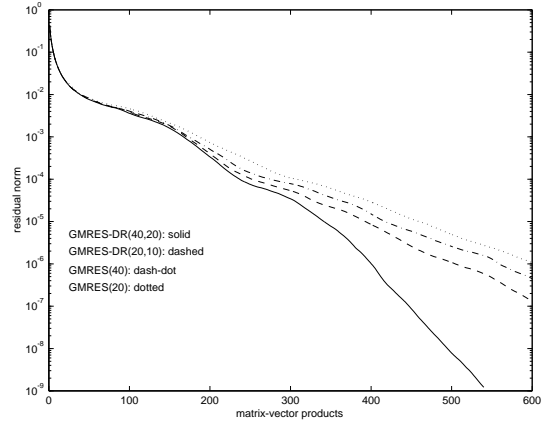}
\vskip-.1cm
\caption{Solution of First Right-hand Side\label{figure1}}
\end{center}
\vskip-.5cm
\end{figure}

{\it Example 1.}  For the first right-hand side, we compare GMRES-DR with
standard GMRES. Figure~\ref{figure1} gives a plot of the residual norms for
GMRES-DR(40,20), GMRES-DR(20,10), GMRES(40), and GMRES(20). 
GMRES-DR(40,20) uses subspaces of dimension 40, of which 20 basis vectors
are approximate eigenvectors. GMRES(40) also restarts when the subspace
reaches dimension 40. Starting around iteration 300, GMRES-DR(40,20)
performs considerably better than the other methods. At that point,
it has developed good enough approximations to some of the smallest 
eigenvalues. The eigenvalues halfway surround the origin~\cite{Medeke}, 
on the positive real side. This makes the problem difficult, but 
removing some of the surrounding eigenvalues nearest the origin is helpful.

\begin{figure}
\begin{center}
\leavevmode
\epsfxsize=2.8 in
\epsfbox{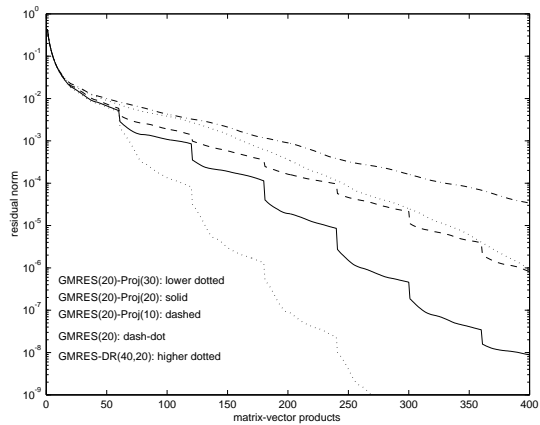}
\vskip-.1cm
\caption{Solution of Second Right-hand Side\label{figure2}}
\end{center}
\vskip-.5cm
\end{figure}

{\it Example 2.}  For the second right-hand side, we compare GMRES-Proj
with standard GMRES and also with GMRES-DR. With GMRES, there is no
deflation, and with GMRES-DR, the deflation happens only after accurate
enough approximate eigenvectors develop. Meanwhile, GMRES-Proj is able to
deflate from the start. Figure~\ref{figure2} has the residual norms for
GMRES(20)-Proj(30), GMRES(20)-Proj(20), GMRES(20)-Proj(10), GMRES(20), and
GMRES-DR(40,20). GMRES(20)-Proj(30) refers to cycles of GMRES(20) with
occational projections over 30 approximate eigenvectors in between. We
project in between every third cycle. GMRES(20)-Proj(20) and
GMRES(20)-Proj(10) use the approximate eigenvectors generated while solving
the first right-hand side with 400 iterations of GMRES-DR(40,20) and
GMRES-DR(20,10), respectively. GMRES(20)-Proj(30)'s eigenvectors come from 
610 iterations of GMRES-DR(50,30) (more iterations are needed in this
case for the eigenvectors to become accurate enough). The expense per 
iteration is nearly the same for GMRES(20) and all the GMRES-Proj methods, 
while GMRES-DR(40,20) has greater orthogonalization expense. More storage is
needed for the GMRES-Proj methods than for GMRES. For example,
GMRES(20)-Proj(20) needs over 40 vectors of length $n$, while GMRES(20) uses
a little over 20. The deflation in GMRES-Proj makes it much better than
GMRES(20). GMRES(20)-Proj(20) is also considerably better than 
GMRES-DR(40,20), because the deflation can start from the beginning 
(both use 20 Krylov vectors and 20 eigenvectors for each cycle). 

\section{Conclusion}  

Deflating eigenvalues is useful for lattice QCD problems, particularly for 
the second and subsequent right-hand sides. While Z(2) noise vectors were
used for these examples, the speedup from deflation is independent of the
nature of the right-hand side.
Future plans include implementing multiple mass versions of GMRES-DR and
GMRES-Proj that solve several systems of equations with shifted matrices 
but the same right-hand side. We also would like to investigate deflating 
eigenvalues for the subsequent right-hand sides from Lanczos methods such 
as the conjugate gradient method and BiCGSTAB.

\section{Acknowledgements}

The work of both authors is partly supported by the Baylor University 
Sabbatical Program and NCSA and utilized the SGI Origin 2000 System at the 
University of Illinois. WW is also partly supported by NSF 
Grant No.\ 0070836.

\end{document}